\documentclass[9pt, one column, revtex4-2]{article}	

\usepackage{geometry}                		
\usepackage{graphicx}	
\usepackage{amssymb}
\usepackage{authblk}
\usepackage{hyperref}
\usepackage{caption}
\usepackage{subcaption}
\usepackage{float}

\title{Does Dynamical Wormhole Evolve From Emergent Scenario? }
\author[1]{Dhritimalya Roy\thanks {rdhritimalya@gmail.com}}
\author[1]{Ayanendu Dutta\thanks {ayanendudutta@gmail.com}}
\author[2]{Bikram Ghosh\thanks {bikramghosh13@gmail.com}}
\author[3]{Subenoy Chakraborty\thanks {schakraborty.math@gmail.com}}

\affil[1]{Department of Physics, Jadavpur University, Kolkata-700032, INDIA}
\affil[2]{Department of Mathematics, Naba Barrackpore Prafulla Chandra Mahavidyalaya, Kolkata-700131, INDIA}
\affil[3]{Department of Mathematics, Jadavpur University, Kolkata-700032, INDIA}

\date{}							

\begin{document}
	\maketitle
	
	\begin{abstract}
		In the present work we analyse a dynamical wormhole solution with two fluids system (one isotropic and homogeneous and the other being inhomogeneous and anisotropic in nature)  as the matter at the throat. We choose two different forms of  Equation of State(EoS) and investigate two solutions of the wormhole geometry. The properties to ensure existence and traversability has been analysed. Also, the model of the dynamic wormhole has been examined for a possibility of the Emergent Universe(EU) model in cosmological context. Finally, for the dynamical wormholes so obtained, Null Energy Condition(NEC) has been examined near the throat. 
	\end{abstract}
	
	Keywords: Dynamical Wormhole; Evolving wormhole; Emergent universe; Energy condition 
	
	\section{Introduction}
	
	Wormholes  are interconnecting bridges between two asymptotic points of the same universe or between two distant universes. The concept of these interconnecting bridges was first put forward by Einstein and Rosen{\cite{Einstein:1935tc}} and was called the Einstein-Rosen bridge. The term wormhole was later coined by Wheeler and Misner\cite{Misner:1957mt} in 1957. The solution of first-ever Traversable Wormhole was put forward by Morris and Thorne\cite{Morris:1988cz} which opened a new area of Wormhole physics. They concluded that for a Wormhole to maintain traversability the matter contained in the neighborhood of the throat, has to violate the Null energy condition (NEC)-the weakest of all energy conditions. Since classical matter obeys the classical energy conditions, it requires a special type of matter called exotic matter which can violate the NEC. The common example of a wormhole solution is the Ellis wormhole of general relativity \cite{Ellis:1973yv, Ellis:1979bh, Visser:1995cc}. This requirement of exotic matter in the development of the Wormhole solution finds ways to limit the use of exotic matter. This was done by providing Wormhole  solutions in modified theories of gravity or by considering matter that fulfill known energy conditions \cite{Lobo:2009ip, Mishra:2021xfl, Moraes:2017mir, Yousaf:2017hjh,Elizalde:2018frj, Dutta:2023wfg,Chakraborty:2007zi,Forghani:2019zbb}. It is noteworthy that explicit static solutions for wormholes, which adhere to energy conditions throughout spacetime, exist in Einstein-Gauss-Bonnet gravity \cite{Maeda:2008nz}. Moreover, in higher dimensions, certain Lovelock theories introduce terms with higher curvature powers, preventing the violation of energy conditions even in a vacuum where there is an absence of any stress-energy tensor across the entire spacetime\cite{Dotti}.
	
	The idea of a possible mode of quick transport between two very distant stellar  objects through a traversable wormhole  is quite intriguing. Although for this to be achievable, we require polished and advanced physics and mathematics that can overcome the problems of cosmic censorship and causality violation principle.

	All of the referred works above are solutions of static wormhole configurations. Non-static configurations can be obtained by introducing a time dependent metric, giving a dynamic wormhole configuration. Visser\cite{Visser:1989kg}  developed a class of traversable WH both static and dynamic type where he used the technique of cut and paste(with suitable Junction convergence criteria) . Dynamic WH in higher order $R^2$ gravity was investigated by Hochberg\cite{Hochberg:1990is}. The Horizon problem was  addressed by other authors\cite{Hochberg:1992du} by using surgical modified dynamic Wormhole solutions in the early inflationary universe. Numerous solutions on similar lines were done by various authors (one may refer to for detailed reading\cite{Roman:1992xj,Morris:1988tu}). The non-static kind of  Lorentzian WH was put forward by Kar\cite{Kar:1994tz,Kar:1995ss}, and they showed that it is possible to obtain non-static wormhole geometry with throat matter satisfying the energy conditions. The authors showed that it is possible to have a dynamic wormhole in FRW spacetime with all energy conditions satisfied. Later, Hochberg and Visser\cite{Hochberg:1998ii} and Hayward\cite{Hayward:1998pp}  showed that the throat can be considered as an anti trapped surface with violation of NEC as the generic property. Evolving wormholes has also been studied in higher order gravity theories as well as in the context of modified theories of gravity by many authors\cite{Naz:2023pfl,Zubair:2022jjm,Saiedi:2021hkh,Zubair:2020uyb,Bhattacharya:2015oma, Mehdizadeh:2021kgv,Bhattacharya:2021frx,KordZangeneh:2020ixt,KordZangeneh:2020jio,Lobo:2020kxn,Mehdizadeh:2017dhb,Mehdizadeh:2012zz,SB,Yousaf:2019uky}.
	
	The equation of state chosen to describe the matter present in the throat of a wormhole has a great importance in the stability and existence of a dynamic wormhole solution. Phantom matter supported dynamic wormhole  solutions were extensively studied by Cataldo\textit{et.al} in\cite{Cataldo:2008pm,Cataldo:2008ku,Cataldo:2012pw,Cataldo:2013ala}  and  established that dynamic wormhole could be supported by fluid mixture of more than one kind such that the total matter satisfies the energy conditions. Pan {\it et.al} \cite{Pan:2014oaa} studied two fluid supported Dynamic Wormhole (DY.WH) in connection with particle creation mechanism. Maeda \textit{et.al} in\cite{Maeda:2009tk} studied a cosmological wormhole satisfying NEC and connected two FRW universities with the whole space time being trapped and unlike only the throat being trapped.

	The standard big bang cosmology has a problem with horizon and singularity like the event horizon of a black hole. It is of interest to look for cosmological models which are free from these pathologies.  Ellis and  Marteens\cite{Ellis:2002we} proposed such a cosmological model which solves the problem of singularity within the realm of Einstein's general relativity, and is known as Emergent Universe(EU). It solves the long-standing problem of singularity. They proposed a nonsingular cosmology which originates as Einstein's static universe and also gets rid of horizon problems. Since then it has gathered a lot of attention among cosmologists and numerous studies have been made in this field\cite{Shekh:2023tbh,Sengupta:2023ysx,Shekh:2023rea,Ellis:2003qz}. Mukherjee\textit{et.al} in their  paper\cite{Mukherjee:2005zt,Mukherjee:2006ds} extended the emergent universe scenario for a flat space time and also obtained the emergent universe solution within a relativistic context. The equation of state considered accommodates the scope for exotic matter which violates the energy conditions of  General Relativity and also accommodates the present late time acceleration of the universe. Several authors made their contribution  in the extension of the Emergent Universe model,namely, Chakraborty\cite{Chakraborty:2014ora} studied particle creation as a consequence of emergent universe, Bose\textit{et.al}\cite{Bose:2020xml} investigated Emergent scenario in Hořava–Lifshitz gravity. A model of emergent universe in inhomogeneous spacetime was investigated by Bhattacharya and Chakraborty\cite{Bhattacharya:2016env} and they inferred that it resembles  features of steady state theory of Fred Hoyle\textit{et.al}.
	Dutta\textit{et.al}\cite{Dutta:2016kkl} in their paper extended the EoS by Mukherjee to a more general form by investigating modified chaplygin gas in emergent universe scenario with certain conditions on the parameter, for the EU model. 
	
	In our present work, motivated from the above studies, we intend to investigate dynamic wormhole
	supported with two fluid  systems (one homogeneous and isotropic, other  being inhomogeneous and anisotropic) and the Emergent Universe scenario. Hence, it would be of interest to see if the Equation of State (EoS) chosen for the fluid system exhibits an Emergent Universe scenario possibility or not.
	As Wormholes are spacetimes  free of singularity and horizons, it is interesting to look for their existence in a cosmological model which also prohibits these pathologies. 
	
	The plan of the paper is the following: we begin with an overview of the  basic equations in inhomogeneous FLRW model with two non-interacting fluids in section\ref{s2} and thereafter possible wormhole(WH) solutions using two choices of the equation of state are presented in section{\ref{s3}}. Further in section\ref{s4} the required conditions  for the validity of the obtained  shape functions are analyzed for both the choices of the equation of state. In the following sections \ref{EU} and \ref{EC} a possibility of an emergent scenario is investigated in Dynamical wormhole spacetime and  the energy conditions are analyzed respectively. 
	At the end in section\ref{con} a discussion about the results obtained in the work has been presented.

	\section{Basic Equations in Inhomogeneous FLRW model: An Overview}\label{s2}
	The line element for inhomogeneous FLRW model is given by:
	
	\begin{equation} \label{eqn1}
		ds^2 = -e^{2\phi(r,t)}dt^2 + a^2(t) \left [\frac{dr^2}{1- \frac{b(r)}{r} - Kr^2} + r^2 d{\Omega_2 }^2  \right],
	\end{equation}
	where, $ {\phi(r,t)}$ is the redshift function, $a(t)$ is the scale factor of the spacetime geometry, $d{\Omega_2 }^2 = d\theta ^2 + sin^2\theta d\phi^2$ , and $K$ is a constant.
	
	For the matter distribution , two non interacting fluids (namely, Fluid 1 and Fluid 2) are considered - one with dissipative  homogeneous fluid and the other being anisotropic and inhomogeneous. 
	
	The energy momentum tensor for the Fluid 1 has the expression:
	\begin{equation}\label{eqn2}
		T^{I}_{\mu \nu} = (\rho_1 + p_1 + \Gamma) u_\mu u_\nu + (p_1 + \Gamma)g_{\mu \nu},
	\end{equation}
	where, $\rho_1 = \rho_1(t)$ is the energy density , $p_1 = p_1(t)$ is the isotropic pressure , $\Gamma$ is the pressure due to dissipation and $u_{\mu}$ is the four velocity of the fluid.
	Further, for the Fluid 2, the energy momentum tensor  is given by:
	\begin{equation}\label{eqn3}
		T^{II}_{\mu \nu} = (\rho_2 + p_t ) v_\mu v_\nu +p_t g_{\mu \nu}+ (p_r + p_t)\xi_\mu \xi_\nu.
	\end{equation}
	Here, $\rho_2=\rho_2(r,t)$ is the energy density of the inhomogeneous fluid ,$p_r=p_r(r,t)$ is the radial pressure, $p_t=p_t(r,t)$ is the transverse pressure of the anisotropic Fluid 2.  $v_{\mu}$ and $\xi_{\mu}$ are unit timelike and spacelike vector respectively, i.e $v_{\mu}v^{\nu}=\xi_{\mu}\xi^{\nu}=-1$ and $\xi^{\mu}v_{\mu}=0$.
	
	The explicit form of  Einstein's field equations
	\begin{equation}
		G_{{\mu}{\nu}} = -\kappa T_{{\mu}{\nu}},
	\end{equation} 
	for the above spacetime model are given by \cite{Pan:2014oaa}
	
	\begin{equation} \label{eqn5}
		3 e^{-2\phi(r,t)} H^2 + \frac{b'}{a^2 r^2} + \frac{3K}{a^2} =  \kappa \rho_1 + \kappa \rho_2 - \Lambda,
	\end{equation}
	
	\begin{equation}\label{eqn6}
		- e^{-2\phi(r,t)} \left(\frac{2 \ddot{a}}{a} + H^2\right) + \frac{K}{a^2} - \frac{b}{a^2 r^3} + 2 e^{-2\phi(r,t)} H     \frac{\partial \phi }{\partial t} +\frac{2}{a^2 r^2 }(r-b) \frac{\partial \phi }{\partial r} = \kappa (p_1 + \Gamma) + \kappa p_r - \Lambda,
	\end{equation} 
	
	\begin{equation}\label{eqn7}
		e^{-\phi(r,t)} \left(\frac{2 \ddot{a}}{a} + H^2\right) + \frac{K}{a^2} + \frac{b- rb'}{2 a^2 r^3} + 2 e^{-2\phi(r,t)} H  \frac{\partial \phi }{\partial t} + \frac{2 r- b- r b'}{2a^2 r^2 } \frac{\partial \phi }{\partial r}+ \frac{r-b'}{ a^2 r}\left[ \left( \frac{\partial \phi }{\partial r}\right)^2 +\frac{\partial^2 \phi }{\partial^2 r}\right]    = \kappa ( p_1 + \Gamma) + \kappa p_t - \Lambda,
	\end{equation}
	
	\begin{equation}\label{eqn8}
		2 \dot{a} e^{-\phi(r,t)} \left(\sqrt{\frac{r-b(r)}{r}}\right) \frac{\partial \phi(r,t) }{\partial r} = 0,
	\end{equation}
	
	where $\kappa$ = 8$\pi$G , $ H = \frac{\dot{a}}{a}$ is the Hubble parameter, an `overdot' denotes the  differentiation w.r.t time $t$ and `prime'  denotes the differentiation w.r.t. to radial coordinate $r$.
	
	From the field equation (\ref{eqn8}),we have two possibilities, namely: $\dot{a}=0$ (static case)  and $\frac{\partial \phi}{\partial r}=0$(non static case). As in the present work we deal with a non-static spacetime geometry therefore we opt for the 2nd possibility with $\phi=0$ without the loss of generality (re-scaling  of the time coordinate).
	
	Thus the metric simplifies to 
	
	\begin{equation}\label{eqn9}
		ds^2 = - dt^2 + a^2(t) \left [\frac{dr^2}{1- \frac{b(r)}{r} - Kr^2} + r^2 d{\Omega_2 }^2  \right].
	\end{equation}
	Also the above field equations simplify to,
	\begin{equation}\label{eqn10}
		3  H^2 + \frac{b'}{a^2 r^2} + \frac{3K}{a^2} = \kappa \rho_1 - \Lambda,
	\end{equation}
	
	\begin{equation}\label{eqn11}
		-  \left(\frac{2 \ddot{a}}{a} + H^2\right) + \frac{K}{a^2} - \frac{b}{a^2 r^3}  = \kappa (p_1 + \Gamma) + \kappa p_r - \Lambda,
	\end{equation} 
	
	\begin{eqnarray}\label{eqn12}
		- \left(\frac{2 \ddot{a}}{a} + H^2\right) + \frac{K}{a^2} + \frac{b- rb'}{2 a^2 r^3}   = \kappa (p_1 + \Gamma) + \kappa p_t - \Lambda.
	\end{eqnarray}
	
	Since the fluids are non interacting in nature so the conservation equations take the form:
	
	\begin{equation}\label{eqn13}
		\frac{\partial \rho_1}{\partial t} + 3H(\rho_1 + p_1 + \Gamma)=0,
	\end{equation}
	\begin{equation}\label{eqn14}
		\frac{\partial \rho_2}{\partial t} + H(3\rho_2 + p_r + 2p_t)=0,
	\end{equation}
	\begin{equation}\label{eqn15}
		\frac{\partial p_r}{\partial r} = \frac{2}{r}(p_t - p_r ).
	\end{equation}
	
	Here, equations (\ref{eqn14}) and (\ref{eqn15}) are the conservation equations for Fluid 2 and the equation(\ref{eqn13}) represents the conservation equation for Fluid 1. One may notice that equation(\ref{eqn15})  is the relativistic Euler equation.
	One important point to note here is that the anisotropic nature of the Fluid 2 (i.e $p_r \neq p_t$) is essential, otherwise there is no inhomogeneity and the fluids becomes non interacting homogeneous fluids, which is physically uninteresting.

	\section{Possible Wormhole Solutions : Mathematical Derivations} \label{s3}
	In order to solve the field equations given by equations (\ref{eqn10})- (\ref{eqn12}), we assume that the EoS for the anisotropic Fluid 2 is of  barotropic in nature and in the present work we shall consider the following two choices, namely:
	$i) p_r = \alpha \rho_2$ and  $ii) p_r = \alpha {\rho_2}^n$,
	
	where, $\alpha$ and $n(\neq1)$ are arbitrary constants.

	\subsection{Case I : $ p_r = \alpha \rho_2$}\label{C1}
	
	We further assume that the two components of the anisotropic pressure are linearly related {\it i.e}
	\begin{equation}\label{eqn16}
		p_t = \omega_t p_r .
	\end{equation}
	Using this assumption the Euler equation(\ref{eqn15}), has the solution
	\begin{equation}\label{eqn18}
		p_r(r,t) = p_{r0}(t) r ^{2(\omega_t - 1)}
	\end{equation}
	with $p_{r0}(t)$ an arbitrary integration function.
	
	Similarly, using equation(\ref{eqn16}) in equation(\ref{eqn14}) , gives the energy density as:

	\begin{equation}\label{eqn20}
		\rho_2(r,t) =	\rho_0(r) a^{(3+ \alpha(1 + 2\omega_t))}
	\end{equation}
	with $\rho_0(r)$ an arbitrary function.
	
	However due to barotropic EoS one may estimate the above arbitrary(integration) functions in equation(\ref{eqn18}) and (\ref{eqn20})as,
	
	\begin{equation}\label{eqn19}
		p_{r0}(t)= \alpha a^{\mu} ~~~~~~ and ~~~~~~  \rho_0(r) = r ^{2(\omega_t - 1)},
	\end{equation}
	with , $\mu = (3+ \alpha(1 + 2\omega_t))$.
	
	Combining the above relations, we get,
	\begin{equation}\label{eqn23}
		\rho_2(r,t) =  r ^{2(\omega_t - 1)} a^\mu ~~~~~ and ~~~~ p_r(r,t) = \alpha r ^{2(\omega_t - 1)} a^\mu .
	\end{equation}
	
	At this point, after obtaining the values of the energy density and radial pressure for the Fluid 2, we now try to obtain a viable solution for the shape function  using the field equations.
	
	Now to obtain the shape like function $b(r)$ we take the difference between equations (\ref{eqn11}) and (\ref{eqn12}), and using equations(\ref{eqn16}),(\ref{eqn18}), and (\ref{eqn19}) one gets the differential equation in $b$ as,
	\begin{equation}
		\frac{b- rb'}{2  r^3} +  \frac{b}{ r^3} = \kappa  (\omega_t - 1)  \alpha  r ^{2(\omega_t - 1)} a^{\mu + 2} .
	\end{equation}
	Now for consistency, the above equation demands ,  
	\begin{equation}
		\mu + 2 = 0,
	\end{equation}
	
	and the differential equation for $b$ becomes,
	\begin{equation}
		rb' -3b = 2 \kappa \alpha (1 - \omega_t) r ^{(2\omega_t + 1)}
	\end{equation}
	which has a solution,
	\begin{equation}\label{eqn29}
		b(r) = b_0 r^3 - \kappa \alpha r ^{(2\omega_t + 1)},
	\end{equation}
	where $b_0$ is the integration constant.
	
	\subsection{Case II: $p_r = \alpha {\rho_2}^n$ , $n\neq1$}\label{C2}
	
	This choice of EoS together with the relation(\ref{eqn16}), one gets the differential equation for $\rho_2$ from the energy conservation equation(\ref{eqn14}) as,
	
	\begin{equation}\label{eqn31}
		\frac{\partial \rho_2}{\partial t} + H(3 \rho_2 + \alpha{\rho_2}^n(1 + 2\omega_t))=0,
	\end{equation}
	
	which has a solution of the form,
	\begin{equation}\label{eqn33}
		\rho_2^{(n-1)} = 3 \rho_{20}(r) / [a^{9(n-1) }- \alpha \rho_{20}(r) (1+2 \omega_t)].
	\end{equation}
	with $\rho_{20}(r)$ an arbitrary integration function.

	Now using equations (\ref{eqn18}) and (\ref{eqn33}) in EoS({\it i.e} $p_r = \alpha {\rho_2}^n$), one has,
	\begin{equation}\label{eqn34}
		p_{r0}(t) r ^{2(\omega_t - 1)} =  \alpha [3 \rho_{20}(r)]^{(\frac{n}{n-1})} / [a^{9(n-1)} - \alpha_0]^{(\frac{n}{n-1})}.
	\end{equation}
	From the above equation, equating the functions of $r$ and $t$ separately, the arbitrary functions $\rho_{20}(r)$ and $p_{r0}(t)$ can be determined as,
	\begin{equation}\label{28}
		p_{r0}(t) = \alpha / [a^{9(n-1)} - \alpha_0]^{(\frac{n}{n-1})},
	\end{equation}
	and 
	\begin{equation}
		\rho_{20}(r) =\left[\frac{r^{2(\omega_t - 1)}}{3}\right]^{(\frac{n-1}{n})} ,
	\end{equation}
	with , $\alpha_0=\alpha \rho_{20} (1+2 \omega_t)$.
	Thus,
	\begin{equation}\label{eqn37}
		\rho_2(r,t) = 3^{[\frac{1}{n (n-1)}]} r^{\frac{2(\omega_t-1) }{n}} / [a^{9(n-1)} - \alpha_0]^{(\frac{1}{n-1})},
	\end{equation}
	and,
	\begin{equation}\label{eqn38}
		p_r (r,t) =3^{[\frac{1}{(n-1)}]} \alpha r^{2(\omega_t - 1)}/ [a^{9(n-1)} - \alpha_0]^{(\frac{n}{n-1})}.
	\end{equation}
	
	Then, similar to the previous case subtracting equation(\ref{eqn11}) from equation(\ref{eqn12}), one gets:
	
	\begin{equation}
		b - rb' + 2b = [\kappa 2 a^2 r^3 (\omega_t -1) \alpha r^{2(\omega_t - 1)}] / [a^{9(n-1)} - \alpha_0]^{(\frac{n}{n-1})},
	\end{equation}
	where relation{(\ref{eqn16})} and the expression for $p_r$ from equations({\ref{eqn18}}) and ({\ref{28}}) has been used. Now 
	for consistency of the above differential equation in $'b'$ one should restrict the arbitrary constants as: $\alpha_0 = 0$ {\it i.e. , } $\omega_t = - 1/2$ and  $n = 2/9$. As a result  differential equation for $``b"$ simplifies to , 
	\begin{equation}
		rb' - 3b = 3 \kappa \alpha,
	\end{equation}
	
	having solution,
	\begin{equation}
		b(r) = b_1r^3 - \kappa \alpha,
	\end{equation}
	where, $b_1$ is the constant of integration.

	It is important to mention that these two solutions of the shape function so obtained must satisfy the flaring out condition for the wormhole to be traversable. Hence in the following section we analyze the condition required for these shape functions to satisfy the flaring out condition.

	\section{An investigation of WH Geometry:  Flaring Out Condition} \label{s4}
	
We shall discuss in this section the conditions imposed by Morris-Thorne for traversability of a WH to check the validity and restrictions on the shape functions obtained in the previous section. The conditions are:
	\begin{itemize}
		\item The throat radius is a global minimum r = $r_0$, so the radial coordinate , $r$ is
		in the interval  [ $r_0$,$\infty$) and the wormhole is constructed by connecting two asymptotic flat regions at the throat.
		\item The redshift function $\phi(r)$ must be finite everywhere in order to avoid the presence of horizons and singularities. So, $e^{\phi(r)} > 0$ everywhere for $r >$$ {r_0}$.
		\item $b(r_0) = r_0$
		and $b^\prime(r)<1 $ at $r=r_0$. Also  for $ r > r_
		0 , b(r) < r$. 
		\item For traversability of the wormhole  the flaring out condition is: 
		$b(r) - r b^\prime(r) > 0 $
		\item Asymptotic flatness:It implies that $\phi(r)$  $\rightarrow$  $0$ and
		$b(r)/r$ $\rightarrow$ $0$ as $r$ $\rightarrow \infty$.
	\end{itemize}

Now the above conditions will be examined for the above two choices of the shape functions obtained in the previous sections.	
	\subsection{Case I}\label{fo1}
The arbitrary constant  $b_0$ in the shape function (\ref{eqn29}) can be obtained from  $b(r_0)=r_0$, ($r_0$ is the throat radius) as,
	\begin{equation}
		b_0 = \frac{[r_0 + \kappa \alpha r_0^{(2\omega_t + 1)}]}{r_0^3},
	\end{equation}
	so the explicit  form of the shape function becomes,
	\begin{equation}\label{eqn46}
		b(r) = \frac{[r_0 + \kappa \alpha r_0^{(2\omega_t + 1)}]}{r_0^3} r^3 - \kappa \alpha r ^{(2\omega_t + 1)}.
	\end{equation}
	The flaring out condition (in the $4^{th}$ bullet) restricts the throat radius as.,

	\begin{equation}\label{res1}
		r_0 \geq \left[\frac{1}{\alpha \kappa (2+\omega_t)}\right]^\frac{1}{2\omega_t},
	\end{equation}
	Thus the shape function in equation(\ref{eqn46}) corresponds to a traversable wormhole provided the throat radius $r_0$ satisfies the  restriction (\ref{res1}). Here the anisotropic fluid, ($\omega_t\neq1$) need not be exotic in nature. Further, the wormhole is infinitely extended over $[r_0, \infty)$.
	
	\subsection{Case II}\label{fo2}
	
	For this solution the throat condition $b(r_0)=r_0$ gives the integration constant $b_1$ as,
	\begin{equation}
		b_1 = \frac{[r_0 + \kappa \alpha ]}{r_0^3}.
	\end{equation}
	Hence, the explicit form of  shape function is,
	\begin{equation}\label{eqn54}
		b(r) = \frac{[r_0 + \kappa \alpha ]}{r_0^3} r^3 - \kappa \alpha.
	\end{equation}
	Now due to the flaring out condition the throat radius $r_0$ is restricted as,
	
	\begin{equation}\label{res2}
		r_0>\frac{\kappa|\alpha|}{2},
	\end{equation}
	provided $\alpha$ is a negative real number.
	
	When this restriction on $r_0$ is obeyed, a proper, viable shape function can be constructed for the given form of EoS. Thus for the second choice of the equation of state it is possible to have traversable wormhole with exotic matter ($\omega_t<-1/2$) of the anisotropic fluid and as in the first choice this wormhole is infinitely extended.
	
	\section{ The Emergent Universe Scenario }\label{EU}
	
	The universe is said to be emergent if it satisfies certain conditions so as to avoid the big bang singularity. The conditions are :
	\begin{itemize}\label{EUCON}
		\item $a\rightarrow a_0$, $H \rightarrow 0$ as $t \rightarrow -\infty$
		\item  $a \simeq a_0$, $H \simeq 0$, $t << t_0$
	\end{itemize}
	
	Inhomogeneous spacetime also exhibits emergent universe scenario which was investigated by Bhattacharya and Chakraborty\cite{Bhattacharya:2016env}. Hence in the present work we examine whether  EU scenario is possible for the above dynamic wormhole configurations
	
	From the line element (\ref{eqn9}) we have,
	\begin{equation}
		ds^2 = - dt^2 + a^2(t) \left [\frac{dr^2}{1 - Kr^2 - \frac{b(r)}{r}} + r^2 d{\Omega_2 }^2  \right],
	\end{equation}	
	
	Using the value of $b(r)$ as obtained in equation(\ref{eqn29}) we get,
	
	\begin{equation}
		ds^2 = - dt^2 + a^2(t) \left [\frac{dr^2}{1 - Kr^2 - b_0 r^2 + \kappa \alpha r^{2 \omega_t}} + r^2 d{\Omega_2 }^2  \right].
	\end{equation}
	
	Here we assume, $K + b_0 = \kappa_0$ and, by proper scaling of the radial coordinate one may have,  $\kappa_0 = 0, \pm1$.
	
	Thus without loss of generality, we have,
	\begin{equation}\label{eqn63}
		ds^2 = - dt^2 + a^2(t) \left [\frac{dr^2}{1 - \kappa_0 r^2 + \kappa \alpha r^{2 \omega_t}} + r^2 d{\Omega_2 }^2  \right].
	\end{equation}
	One important point to note here is that for the present  wormhole solution one has to choose,  $\kappa_0 = 0$, $\alpha<0$, $\omega_t \neq 0,1$. Thus $\alpha<0$ implies, the anisotropic fluid for the first WH solution is also exotic in nature.
	Now, using this value of $b(r)$, $\rho_2$, and $p_r$, $p_t$ from the first case, in the field equations and from the Friedmann equations, we get,
	\begin{equation}\label{eqn64}
		3(H^2 + \frac{\kappa_0}{a^2}) = \kappa \rho_1 - \Lambda,
	\end{equation}
	and,
	\begin{equation}\label{eqn65}
		-(2 \dot{H} + 3 H^2 + \frac{\kappa_0}{a^2}) = \kappa(p_1 + \Gamma) - \Lambda,
	\end{equation}
	with equation(\ref{eqn13}) as the conservation equation. Now, we choose, $p_1 = \omega \rho_1$ and $\Gamma = - (\delta/ \rho_1^n)$, then the conservation equation (\ref{eqn13}) becomes,
	\begin{equation}\label{eqn66}
		\frac{\partial \rho_1}{\partial t} + 3H(\rho_1 + \omega \rho_1 - \delta/ \rho_1^n)=0.
	\end{equation}
	Here the constant $\delta $ may be termed as dissipation coefficient.
	
	The above first order non linear differential equation(\ref{eqn66}) has a solution :
	
	\begin{equation}\label{eqn71}
		\rho_1 = \left(\frac{\delta}{(1+\omega)}+\frac{Z_0 a^{-l}}{(1+\omega)}\right)^{1/(1+n)},~  for ~~ \omega\neq-1
	\end{equation}
	
	and, 
	\begin{equation}\label{eqn75}
		\rho_1=\left(3\delta(1+n)\ln (\frac{a}{a_0})\right)^{(1/[n+1])}, ~for~~ \omega = - 1 ,
	\end{equation}
	
	where, $l=3(1+\omega)(1+n)$ , $Z_0$ and $a_0$ are the integration constants.
	
	In equation (\ref{eqn64}) if , $\kappa_0 = 0$ and  $\Lambda=0$, then the first Friedmann equation is,
	\begin{equation}\label{eqn76}
		3 H^2 =  \kappa \rho_1.
	\end{equation}
Now using the above solution for $\rho_1$ in equation(\ref{eqn76}) gives the scale factor as 
	
	\begin{equation}\label{eqn77}
		\sqrt{3}/2 (1+\omega)Z_0^\beta (t-t_0) = a^{\frac{\sqrt{3}}{2}(1+\omega)} \,_2F_1 [\beta,\beta, \beta+1,- \frac{\delta}{Z_0(1+\omega)} a ^{(\frac{3(1+\omega)}{2 \beta})}],   ~~ for~ \omega\neq-1,
	\end{equation}
and,
	\begin{equation}\label{eqn78}
		a = a_0  \exp\left[\Sigma_0  (t-t_0)^{[(n+1)/n]}\right] ,~ ~for~  \omega=-1,
	\end{equation}
		where, $\Sigma_0 = \left(\frac{\sqrt{\kappa}}{\sqrt{3}} (\frac{n}{n+1})\right)^{[\frac{n+1}{n}]} [3\delta(n+1)]^{1/n}$, and $_2F_1$ is the usual hypergeometric function.
	
\begin{figure}[h]
		\centerline{\includegraphics[scale=.7]{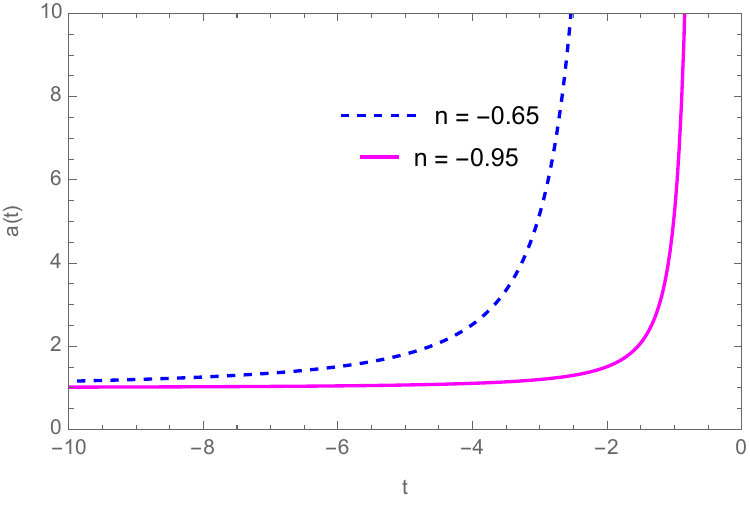}}
\caption{ Graphical representation of $'a(t)'$ in equation (\ref{eqn78}) for $\omega=-1$ against  $'t' $  with $\delta = 1$, $a_0=1$, $t_0=0$.}
		\label{plot1}
	\end{figure}

	In the above solution(\ref{eqn78}) for the scale factor if  $-1<n<-1/2$, {\it i.e} $\beta>1$ then $a\rightarrow a_0$ as $t\rightarrow-\infty$. Thus from the above  conditions of Emergent Scenario and the graph of $a(t)$ in fig(\ref{plot1}) ,  shows that the solution (\ref{eqn78}) of the scale factor provides a possible Emergent Universe scenario.
	
	\section{Energy Conditions}\label{EC}
	The formation of traversable wormholes possesses a particularly intriguing and somewhat distinctive characteristic, involving the necessity of exotic matter, which contradicts classical energy conditions. This exotic matter is essential for maintaining the wormhole throat and thereby ensuring its traversability. As outlined in the introduction, extensive research has been conducted on this feature in the context of evolving traversable wormholes and Morris-Thorne type wormholes, explored through various versions of modified theories. Interestingly, some of these theories support forms of matter that meet the energy conditions in diverse ways. Notably, even if standard matter satisfies the conditions, the presence of coupled matter is required to act as exotic energy, fulfilling the traversability requirement. Ref.\cite{Roy:2022kid} provides a detailed examination of this phenomenon, particularly in terms of geometrical matter. The energy conditions namely: Null Energy Condition(NEC), Weak Energy Condition(WEC), Dominant Energy Condition(DEC) and Strong Energy Condition(SEC) are satisfied for two fluid system when the following conditions are met:
	\begin{itemize}
		\item (i)NEC: $\rho_T+{(p_r)}_T \geq 0$, $\rho_T+{(p_t)}_T \geq 0;$
		\item (ii)WEC:$\rho_T \geq 0$,$\rho_T+{(p_r)}_T \geq 0$, $\rho_T+{(p_t)}_T \geq 0$;
		\item (iii)SEC:$\rho_T+{(p_r)}_T \geq 0$, $\rho_T+{(p_t)}_T \geq 0;$ $\rho_T+{(p_r)}_T +2{(p_t)}_T \geq 0$;
		\item (iv)DEC:$\rho_T \geq 0$,$\rho_T-|{(p_r)}_T| \geq 0$, $\rho_T-|{(p_t)}_T| \geq 0$.
	\end{itemize}
	where, $\rho_T = \rho_1+\rho_2$ is the total energy density, ${(p_r)_T} = p_1 + \Gamma + p_r$ is the total radial pressure and ${(p_t)}_T= p_1 +\Gamma + p_t$ is the total transverse pressure for the two fluids combined.

	For the first case the expressions for combined energy conditions for both fluids are:
	
	\begin{equation}
		\rho_T + {(p_r)}_T = {(1+\omega)\rho_1 - \delta \rho_1^{-n} + (1+\alpha)\rho_2},
	\end{equation}
	\begin{equation}
		\rho_T + {(p_t)}_T = {(1+\omega)\rho_1 - \delta \rho_1^{-n} + (1+\omega_t\alpha)\rho_2},
	\end{equation}
	\begin{equation}
		\rho_T + {(p_r)}_T + 2+ {(p_t)}_T = (1+3\omega)\rho_1 - 3\delta \rho_1^{-n}+ [1+\alpha (1+ 2 \omega_t)]\rho_2,
	\end{equation}
	\begin{equation}
		\rho_T - {(p_r)}_T = {(1-\omega)\rho_1 + \delta \rho_1^{-n} + (1-\alpha)\rho_2},
	\end{equation}
	\begin{equation}
		\rho_T - {(p_t)}_T ={(1-\omega)\rho_1 + \delta \rho_1^{-n} + (1-\omega_t\alpha)\rho_2},
	\end{equation}
	
similarly for the second case we get,
\begin{equation}
	\rho_T + {(p_r)}_T = {(1+\omega)\rho_1 - \delta \rho_1^{-n} + (1+\alpha \rho_2^{(n-1)})\rho_2},
\end{equation}
\begin{equation}
	\rho_T + {(p_t)}_T =  {(1+\omega)\rho_1 - \delta \rho_1^{-n} + (1+\omega_t\alpha \rho_2^{(n-1)})\rho_2},
\end{equation}
\begin{equation}
	\rho_T + {(p_r)}_T + 2+ {(p_t)}_T = (1+3\omega)\rho_1 - 3\delta \rho_1^{-n}+ [1+\alpha \rho_2^{(n-1)}(1+ 2 \omega_t)]\rho_2,
\end{equation}
\begin{equation}
	\rho_T - {(p_r)}_T = {(1-\omega)\rho_1 + \delta \rho_1^{-n} + (1-\alpha \rho_2^{(n-1)})\rho_2},
\end{equation}
\begin{equation}
	\rho_T - {(p_t)}_T ={(1-\omega)\rho_1 + \delta \rho_1^{-n} + (1-\omega_t\alpha \rho_2^{(n-1)})\rho_2}.
\end{equation}

	As we are dealing with dynamic wormhole in EU scenario (for which $\omega=-1$),so that the above energy conditions can be written in compact tabular form as follows:

	\begin{table} [h!]
		\centering
		\begin{tabular}{ ||  c c  || }
			\hline
			Energy Conditions &Observations\\ 
			\hline \hline
			$NEC$& Satisfied if , $\alpha\geq \frac{\delta \rho_1^{-n}}{\rho_2} -1= \alpha_1$ and $\omega_t\geq\omega_{t1}=\frac{\alpha_1}{\alpha}$\\
			$DEC$ & Satisfied if , NEC  holds, also $\alpha\leq1+ \frac{\rho_1}{\rho_2}[2+\delta\rho_1^{(-n-1)}]=\alpha_2$ and $\omega_t\leq\omega_{t2}=\frac{\alpha_2}{\alpha}$  \\
			$SEC$ &   Satisfied if, NEC holds along with $\alpha \geq [\frac{\rho_1}{\rho_2}(3 \delta\rho_1^{-(n+1)} +2)-1]/ (1+2\omega_t)=\alpha_3$    \\
			$WEC$ &Satisfied if , condition of NEC holds along with $\rho_T\geq0$\\ [1ex]
			\hline 
		\end{tabular}
		\caption{Range of $\alpha$, $\omega_t$ considering $\omega=-1$ for the Energy conditions to be satisfied for  case I.}
		\label {table:1}
	\end{table}

		\begin{table} [h!]
		\centering
		\begin{tabular}{ ||  c c || }
			\hline
			Energy Conditions &Observations\\ 
			\hline \hline
			$NEC$& Satisfied if , $\alpha\geq\frac{\alpha_1}{\rho_2^{(n-1)}} =\alpha_4$ and $\omega_t\geq\omega_{t3}=\frac{\alpha_4}{\alpha}$   \\
			$DEC$ & Satisfied if , NEC is obeyed, also $\alpha\leq\frac{\alpha_2}{\rho_2^{(n-1)}}=\alpha_5$ and $\omega_t\leq\omega_{t4}=\frac{\alpha_5}{\alpha}$    \\
			$SEC$ &Satisfied if, NEC holds along with $\alpha \geq\frac{\alpha_3}{\rho_2^{(n-1)}}=\alpha_6$    \\
			$WEC$ &Satisfied if, condition of NEC holds along with $\rho_T\geq0$   \\ [1ex]
			\hline 
		\end{tabular}
		\caption{Range of $\alpha$, $\omega_t$ considering $\omega=-1$ for the Energy conditions to be satisfied for case II.}
		\label {table:2}
	\end{table}
	
\begin{figure*}[h!]
		\centering
     	\subfloat[]{{\includegraphics[width=10cm]{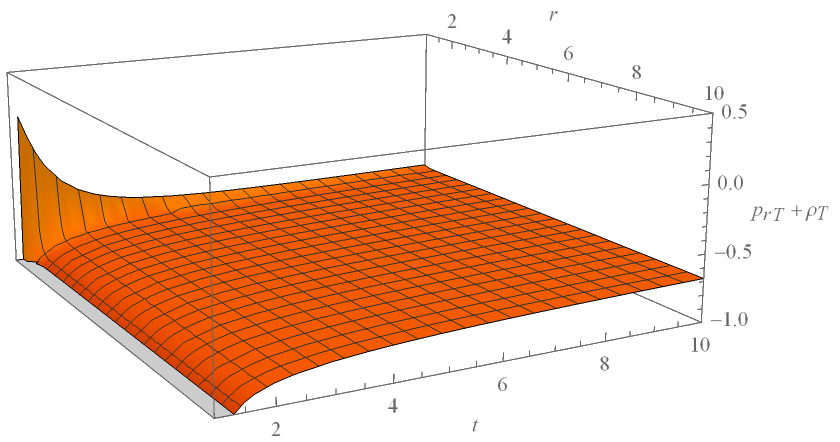}}}\qquad
    	\subfloat[]{{\includegraphics[width=10cm]{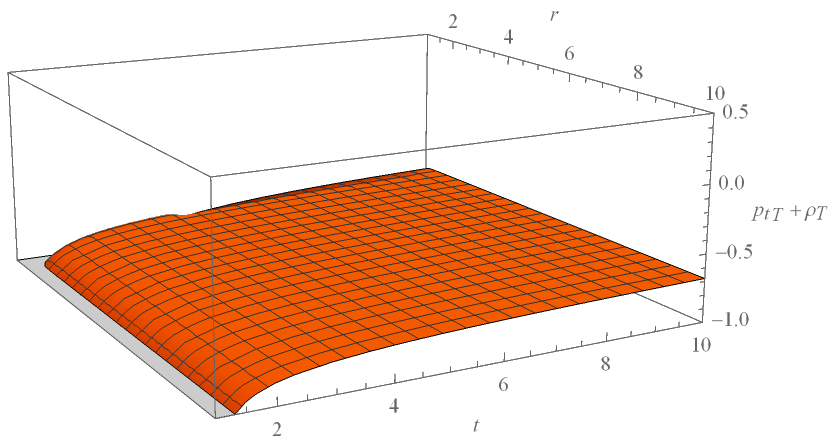}}}\\
		
	\caption{Plots demonstrating the variation of NEC( $ \rho_T+(p_r)_T, ~ \rho_T+(p_t)_T, $) with radial distance $ r $ and scale factor $a$ for Case I. The nature of the plots are obtained considering $ \alpha=0.7,~a_0=1,~n=2/9~\delta=1,~\omega=-1~ $.}
		\label{ec1}
	\end{figure*}

\begin{figure*}[h!]
		\centering
		\subfloat[]{{\includegraphics[width=8cm]{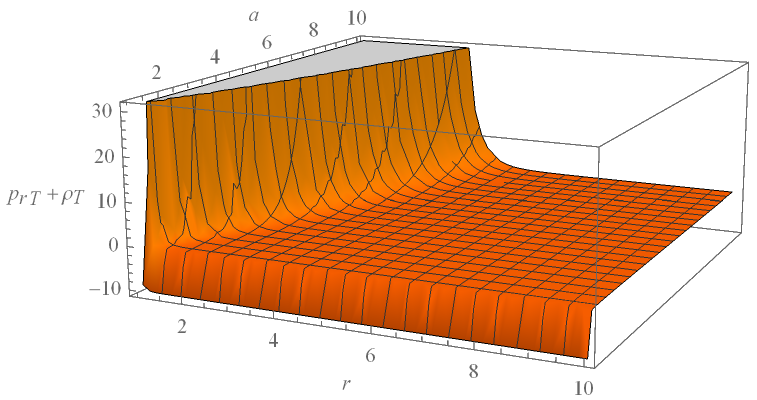}}}
		\subfloat[]{{\includegraphics[width=8cm]{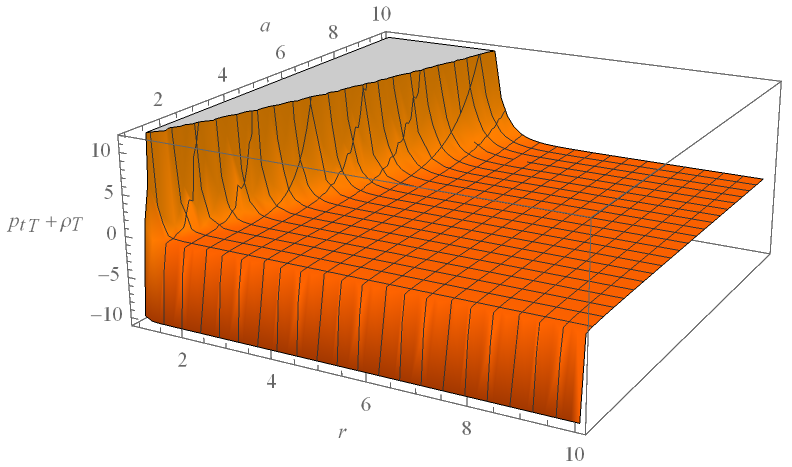}}}\\
		
			\subfloat[]{{\includegraphics[width=8cm]{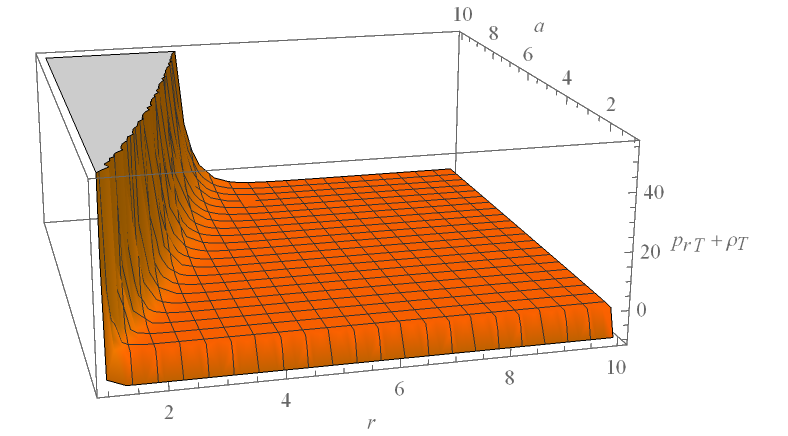}}}	\subfloat[]{{\includegraphics[width=8cm]{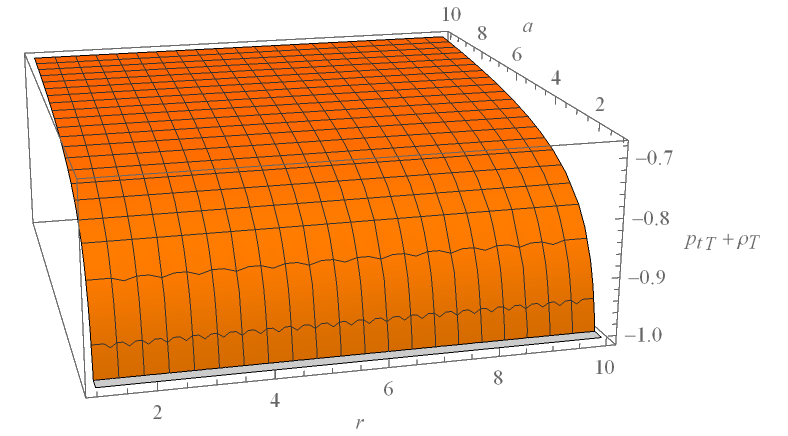}}}\qquad
		\caption{Plots demonstrating the variation of NEC($ \rho_T+(p_r)_T, ~ \rho_T+(p_t)_T, $) with radial distance $ r $ and scale factor $a$ for Case II. The nature of the plots are obtained considering $ \alpha=0.7,~a_0=1,~n=2/9,~\delta=1~\omega_t=-0.5,~ \omega=-1$ for plots ($a$) and ($b$). For plots ($c$) and ($d$), $\alpha=2,~a_0=1,~n=2/9~\delta=1,~\omega_t=-0.5,~ \omega=-1$ }
		\label{ec2}
	\end{figure*}

	Further, to have a clear picture both the conditions for NEC has been plotted graphically in figures (\ref{ec1}) and (\ref{ec2}). For the case I wormhole, $\rho_T + {(p_r)}_T$ is positive very close to the throat while $\rho_T + {(p_t)}_T$ is negative, and this is true for possitive and negative values of $\alpha$. On the other hand for Case II wormhole both the NECs are satisfied for $\alpha<2$ but $\rho_T + {(p_t)}_T$ is violated for $\alpha\geq2$.
	
	
%
%

	\section{Summary} \label{con}
	
	The present work deals with inhomogeneous FLRW spacetime to examine whether  dynamical wormhole solution is possible or not for the non-interacting two fluids system as the matter content. The first fluid(termed as Fluid 1) is homogeneous but dissipative in nature while the other fluid({\it i,e} Fluid 2) is both inhomogeneous and anisotropic in nature. Assuming barotropic equation of state of the form $p_r = \alpha \rho_2$ in the first case while a polytropic equation of state : $p_r = \alpha \rho_2^n$ in the second case, it is possible to have Wormhole solutions with linear relationship between the radial and tangential pressures(the proportionality constant is chosen to be different from unit due to anisotropic nature of Fluid 2). For Fluid 1 a simple barotropic equation of state is assumed , while the dissipative pressure is chosen as power law to the matter density. Both the wormhole solutions has been examined for the throat condition and the traversability condition. As a result, the arbitrary integration constants are estimated as a function of the throat radius and the parameters involved are restricted due to flare-out condition. Subsequently, the scale factor has been evaluated by solving the first Friedmann equation and one has two distinct solutions for $\omega\neq-1$ and $\omega=-1$, where $\omega$ is the barotropic constant for the first fluid. From the asymptotic analysis it is found that the scale factor (for $\omega=-1$ )  behaves as an emergent scenario at infinite past with specific restrictions on the parameters involved. The NEC has been examined graphically and a detailed discussion has been presented. Thus the present study shows that it is possible to have dynamical wormhole with emergent phase at very early cosmic era. Finally, for the present wormhole configuration it is possible to have non-violation of NEC near the throat - a distinct feature for dynamical wormhole compared to the static wormhole solutions. 
	
	Finally, from a cosmological point of view this dynamical wormhole configuration related to emergent scenario may be a model of the early evolution of the universe avoiding the big-bang singularity.

	\section{Acknowledgement}
	S.C. thanks FIST program of DST, Department of Mathematics, JU (SR/FST/MS-II/2021/101(C)).

\end{document}